# In-situ Time-domain Physical Adjoint Optimization of Complex Wave Dynamics

Hoyeong Kwon[1], Arunn Suntharalingam[1], Laureano Bulus-Rossini[2], Tsampikos Kottos[1]*

[1] Wave Transport in Complex Systems Lab, Department of Physics, Wesleyan University, Middletown, Connecticut 06457, USA

[2] Departamento de Ingeniería en Telecomunicaciones - Instituto Balseiro (UNCuyo-CNEA) & CONICET, CCT Patagonia Norte, Bariloche 8400 (RN), Argentina

## ABSTRACT

Direct optimization of complex wave dynamics through the intrinsic evolution of physical systems is fundamentally limited by the lack of directly accessible gradients[1-7]. Adjoint methods provide an exact route to gradient computation[8-11] and have enabled optimization in numerical solvers and, more recently, in frequency-domain physical platforms[12,13]. Yet their extension to the time domain has remained out of reach, as reproducing time-reversed propagation appears to require non-causal operations or compensating gain[14-17]. Here we develop a protocol and experimentally demonstrate that time-domain adjoint dynamics can be realized in-situ in linear physical systems without gain, non-causal elements, or auxiliary backward networks. By combining time remapping with a transformation of system variables, we obtain a physically realizable adjoint evolution that constructs gradients from measurable signals. We experimentally demonstrate the approach in a complex RLC network, realizing in-situ optimization for time-dependent objectives, including time-windowed and broadband responses. This framework unifies physical optimization by enabling both system parameters and source excitations to be optimized within the same platform. Our results close the gap between adjoint theory and physical implementation, establishing a general foundation for hardware-native, in-situ temporal optimization across a broad class of complex dynamical systems.

## INTRODUCTION

Physical systems naturally perform rich transformations through their intrinsic dynamics, converting inputs into structured responses across space, time, and frequency. If such transformations could be optimized directly within the physical platform, it would become possible to shape signals, control wave propagation, and design system responses without relying exclusively on external numerical models. This possibility is particularly attractive in complex dynamical settings, where the physical system itself already embodies the underlying operator and may offer a route to efficient hardware native optimization[1,2].

In practice, however, exploiting physical dynamics for optimization requires access to gradients, which remains a fundamental bottleneck. Whereas digital platforms can compute gradients from an explicit numerical model, physical platforms generally do not provide direct access to derivatives or adjoint propagation[3-7]. As a result, optimization is typically performed using model-based, gradient evaluation schemes[18-22] or gradient-free approaches[23-25]. In both cases, sensitivities are inferred rather than measured directly from the physical system. This reliance limits real time optimization and the direct design of physical responses through intrinsic dynamics. This limitation becomes more severe in complex wave-chaotic systems, where multiple scattering and the interference of numerous propagation paths produce intricate speckle-like fields with strong sensitivities to perturbations; disorder, and fabrication imperfections can therefore introduce mismatch between numerical prediction and physical behavior[19].

*Corresponding author: tkottos@wesleyan.edu

Many practical waves and network systems, including wireless communication, and acoustic sonar, involve signal propagation through complex environments with strong scattering and interference[25-30]. Fig. 1A describes such settings, where both the internal configuration of the system $H(p)$ and the injected waveform $I(t)$ critically determine the observed response. The complexity of the system dynamics, together with their sensitivity to both the input waveform and system parameters, makes accurate prediction difficult in realistic settings and motivates in-situ optimization that incorporates the realized dynamics of the physical system.

Gradient based optimization of dynamical systems is most naturally expressed through adjoint methods, which provide an efficient way to evaluate the sensitivity of an objective function to states, inputs, and system parameters[8-12]. The same principle underlies reverse-mode differentiation in digital computation[31]. In physical dynamic systems, however, the corresponding adjoint evolution is not automatically available and must itself be constructed or realized. This challenge has limited real time adjoint based optimization, despite its broad success in electromagnetic and photonic inverse design[3,4]. Recent progress has further shown that, in some cases, in-situ adjoint based optimization can be implemented in physical systems in the frequency domain[13]. These studies provide proof of concept demonstrations, particularly in steady state or reciprocal wave settings. However, they do not resolve the more general time domain problem[14, 32-34], where objectives depend on the full transient evolution rather than a single steady state field. In this regime, gradient extraction must reproduce temporal sensitivity across causal dynamics, and the adjoint process can no longer be reduced to a simple frequency-domain argument.

The central obstacle is that time domain adjoint optimization mathematically requires backward propagation in time in transposed system[10]. For a general physical system, this requires a backward-in-time evolution of the system dynamics, which in the original physical variables may demand non-causal operations that are not physically available *within the same platform*. This issue is particularly severe in dissipative or directionally structured systems, where naive time reversal in experiment fails to reproduce the correct adjoint evolution. As a result, although time domain adjoint methods are well established mathematically, their direct physical realization has remained elusive. This implementation barrier has limited the use of time-domain adjoint methods as experimental optimization tools and left open whether gradients can be measured in-situ rather than inferred externally.

In this study, we show that time domain adjoint dynamics can be realized in-situ in general linear physical systems without gain, non-causal operations, or auxiliary backward hardware. Our approach reformulates the adjoint dynamics through a reparameterization of time axis and a transformation of system variables. In this way, the system itself generates the signals required for gradient evaluation, enabling exact gradients with respect to the realized physical system to be obtained from measurable responses. The resulting framework establishes a unified route to physical optimization in which both system parameters and sources can be optimized respectively within the same experimental platform. We extend the framework beyond steady-state or single-frequency objectives to general time-dependent tasks, including time-windowed responses, broadband objectives, and temporal demultiplexing. Because gradients are obtained from measurable responses rather than idealized models, the optimization naturally incorporates disorder, loss, and other nonidealities from experiments. This is particularly valuable in complex or imperfect physical systems, where it helps close the gap between numerical design and experimental performance. More broadly, our results establish a general paradigm for hardware-native optimization in which gradients are evaluated directly by physical dynamics.

## Theoretical Background: In-situ Time-Domain Adjoint Optimization Protocol

We consider a linear dynamical system whose evolution is governed by a set of coupled equations describing the forward dynamics of a physical network. In our implementation, the complex environment illustrated in Fig. 1A is emulated by a reconfigurable RF network of coupled RLC resonators (Fig. 1B), whose state variables correspond to directly measurable voltages or currents. The in-situ time-domain adjoint optimization protocol consists of a forward experiment, an adjoint experiment *performed on the same physical platform*, and a gradient evaluation procedure. By combining measured forward and adjoint responses, the framework enables gradients with respect to both system parameters and source excitations to be obtained directly from experiment. For clarity, we present below the theoretical foundation of the protocol for the specific case of RLC networks, while a platform-agnostic general derivation is provided in Method and Supplementary Information Section 1.1[35].

The forward dynamics of the physical system is derived using Kirchhoff's Law[36] and can be described in a compact form as

$$F: \dot{\psi}(t) + H(\boldsymbol{p})\psi(t) - s(t) = 0, \tag{1}$$

where $\psi(t) = \left(v_1(t), \ldots, v_n(t), i_1(t), \ldots, i_n(t)\right)^T$ denotes the system variables vector of voltage $v_l(t)$ and inductance current $i_l(t)$ at each node $l = 1, \ldots, n$ and $H(\boldsymbol{p})$ is the dynamical operator determined by the physical parameters such as coupling elements, and resistance and reactance of each RLC node which are collectively indicated with the vector $\boldsymbol{p}$ . In our experiment, the coupling is realized with tunable inductance. The term $s(t)$ is the forward source excitation.

The objective functional $G\left(\psi, \boldsymbol{p}, s(t)\right) = \int_0^T g(\psi, \boldsymbol{p}, s(t), t)dt$ is defined from the measurable forward response and may depend on the full temporal evolution of the system over a prescribed observation window. To evaluate gradients of the objective function efficiently, we introduce an adjoint methodology. In this framework, the sensitivity of system parameter optimization is derived as

$$\frac{dG}{d\boldsymbol{p}} = -\int_0^T \left(\phi(t)\right)^T \frac{\partial H(\boldsymbol{p})}{\partial \boldsymbol{p}} \psi(t)\, dt, \tag{2}$$

where $\phi(t)$ is the Lagrange multiplier that is evaluated as a solution of the adjoint equation. The corresponding adjoint equation is derived rigorously from the standard variational formulation that evolves backward in time and involves the transpose of the system operator ($H^T(\boldsymbol{p})$), i.e.,

$$A: -\dot{\phi}(t) + H^T(\boldsymbol{p})\phi(t) - \left(\frac{\partial g}{\partial \psi}\right)^T = 0. \tag{3}$$

where $s_{adj} = \left(\frac{\partial g}{\partial \psi}\right)^T$ plays the role of the source for the adjoint problem. Obviously, Eq. (3), differs from Eq. (1) that describes the forward propagation, and therefore, cannot be implemented at the same physical platform[10]. Thus, $\phi(t)$ cannot be extracted by measurements using the same hardware, indicating that an in-situ implementation of the time-domain adjoint optimization, via the evaluation of the sensitivities Eq. (2), is not possible in general.

To recast the adjoint dynamics into a physically realizable form that can be executed within the same hardware used for the forward dynamics given by Eq. (1), we first introduce a change of variables defined on a finite time interval $T$ as $\tau = T - t$, which remaps the backward-time adjoint equation into an equivalent forward-time formulation (see Supplementary Information Section 1.2.1) [35]. Under this reparameterization, the adjoint dynamics are expressed on the same forward time axis as the physical system, so that both the evolution and the associated time integration are carried out along the forward temporal direction in the lab. Importantly, this transformation is a mathematical reparameterization of the time variable and does not correspond to a physical time reversal of signals in the experiment[15]. We then identify a variable transformation $\Theta$ satisfying $H = \Theta H^T \Theta^{-1}$, which converts the adjoint operator into the same dynamical operator governing the forward system Eq. (1) (see Supplementary Information Section 1.2.2 [35]) with a time-domain solution $\tilde{\phi}(t) = \Theta\phi(t)$ being a directly measurable field generated by our hardware. As a result, both forward and adjoint dynamics can be realized on the same physical hardware while evolving on the same forward time axis. Then, Eq. (2) can be recast in terms of the experimentally measured "adjoint" response $\tilde{\phi}(t)$ as

$$\frac{dG}{d\boldsymbol{p}} = -\int_0^T \left(\Theta^{-1}\tilde{\phi}(t)\right)^T \frac{\partial H(\boldsymbol{p})}{\partial \boldsymbol{p}} \psi(t)\, dt, \tag{4}$$

It is important to emphasize that Eqs. (2-4) do not require measuring the entire $\tilde{\phi}(t)$ or $\psi(t)$ but only those components that correspond to the non-zero entries of $\frac{\partial H(\boldsymbol{p})}{\partial \boldsymbol{p}}$ associated with the controllable parameters. In other words, the sparsity of $\frac{\partial H(\boldsymbol{p})}{\partial \boldsymbol{p}}$ reduces measurement complexities dramatically. Let us also point out that the evaluation of $\Theta^{-1}$ and its multiplication with the state-vector $\tilde{\phi}(t)$ are inexpensive computational operations since $\Theta$ is a sparse symmetric block matrix.

In source optimization, the sensitivity can also be evaluated and is defined locally in time as

$$\frac{dG}{ds(t)} = \phi(t) = \Theta^{-1}\tilde{\phi}(t), \tag{5}$$

which introduces a causality constraint. As the optimization window occurs after the source excitation, a direct application of the adjoint source would require access to future information of the source, violating causality. To enforce a physically realizable excitation, we introduce a time reverse in experiment by flipping the adjoint source as $t \to T - t$, where $T$ is the total duration of the experiment. Under this transformation, the experimental adjoint source is generated as $\tilde{s}_{adj}(T - t)$, and the corresponding adjoint response is measured as $\tilde{\phi}(T - t)$. The measured signal is then mapped back to the original time coordinate $t$ to evaluate the sensitivity. This time reverse in source optimization is distinct from the earlier time reparameterization used in the adjoint derivation. The transformation $\tau = T - t$ is a mathematical reparameterization that enables a forward-time formulation of the adjoint dynamics, whereas *the* $t \to T - t$ *is an experimental procedure introduced only in source optimization* to enforce causality in the generation of the adjoint excitation. A detailed derivation of the full theory is provided in the Supplementary Information Section 1.2 [35].

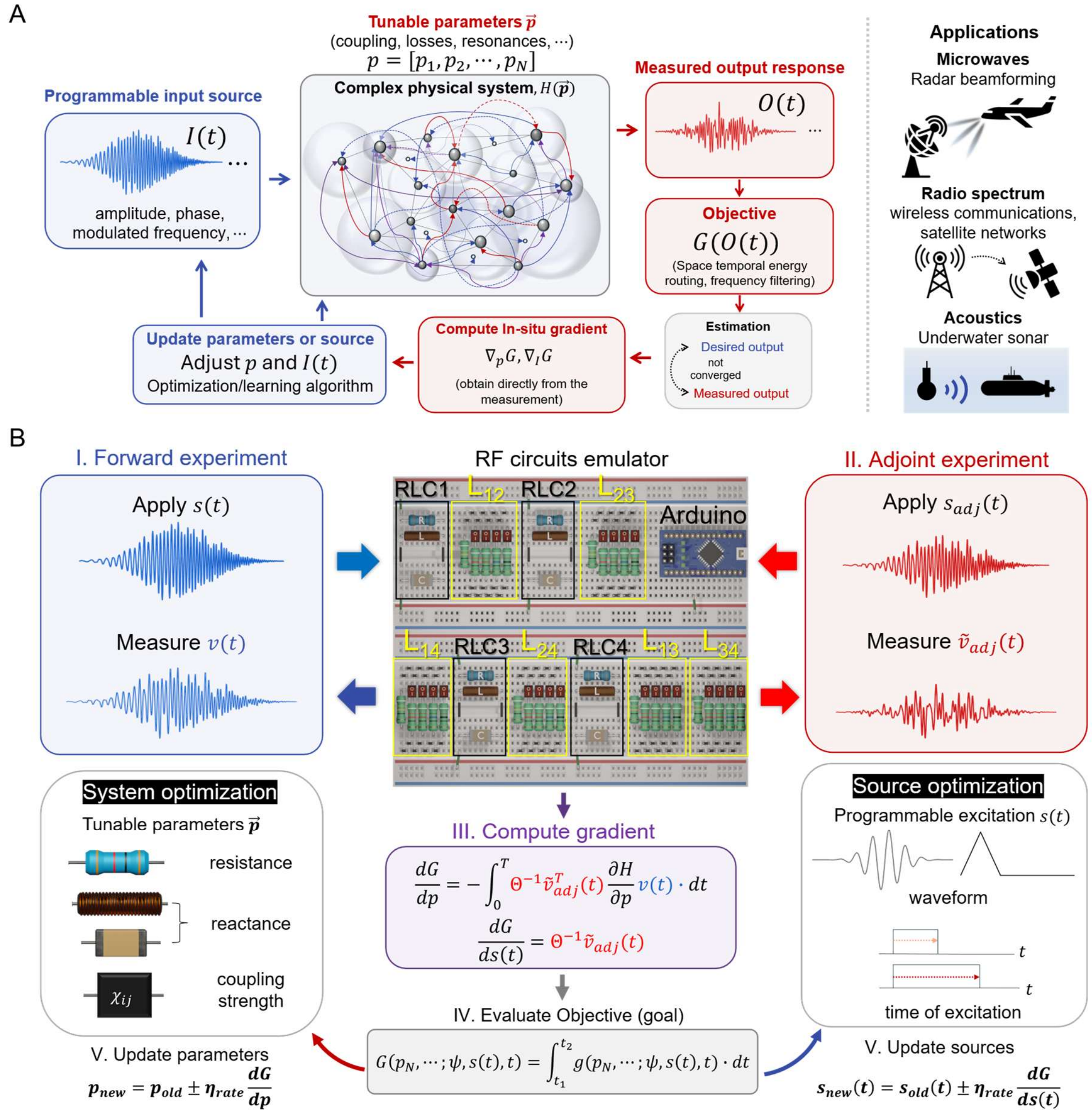

**Fig. 1 | Framework for in-situ adjoint optimization in physical dynamical systems.**
(A) Conceptual framework for hardware-native optimization. A programmable input waveform $s(t) = I(t)$ excites a complex physical system characterized by tunable parameters $p$, including coupling strengths, losses, and resonances of arbitrary system. The measured response $O(t)$ is evaluated through a user-defined objective function $G$, and gradients with respect to both system parameters and source excitations are obtained directly from the physical system. The resulting sensitivities are used to iteratively update either the system configuration or the input waveform. Representative applications include microwave beamforming, wireless communication, and acoustic sensing. (B) Experimental realization of the in-situ adjoint optimization framework using a reconfigurable RF circuit emulator composed of coupled RLC resonators. The procedure consists of (I) a forward experiment, in which an excitation $s(t)$ is applied and the forward response $v(t)$ is measured; (II) an adjoint experiment, in which an objective-dependent adjoint excitation $s_{adj}(t)$ is applied and the adjoint response $\tilde{v}_{adj}(t)$ is measured; and (III) gradient evaluation using the measured forward and adjoint responses. The transformation $\Theta$ maps the physically realizable response $\tilde{v}_{adj}(t)$ to the adjoint field that is needed for the evaluation of gradients (see text). The experimentally obtained gradients enable optimization of either physical system parameters, such as resistance, reactance, and coupling strength, or programmable source waveforms and excitation timing. Steps (I) to (V) are repeated iteratively until the desired objective is achieved.

Under this construction, the adjoint excitation is generated directly from the objective function and applied experimentally to the same physical platform and time axis. The resulting adjoint response is measured and combined with the forward response to evaluate gradients. This framework naturally supports two classes of optimization: system parameter optimization and source waveform optimization. The detailed derivation of sensitivity is provided in Supplementary Information Section 2.2 and 2.3.

## Experimental Implementation of In-situ Time-Domain Adjoint Optimization

We experimentally demonstrate in-situ parameter and source optimization using a coupled RLC network with a multiply connected topology that supports complex scattering and interference among numerous propagation paths. The platform enables both forward and adjoint measurements on the same hardware, allowing gradients to be extracted directly from measured signals without reconstructing the system model as discussed. First, the experimental implementation of parameter optimization is shown in Fig. 2A. A programmable controller coordinates the arbitrary waveform generator and oscilloscope, while the RLC network emulator of fully coupled tetramer realizes the system dynamics. Forward and adjoint experiments are performed sequentially by switching the measurement configuration, and the resulting signals are used to compute gradients and update circuit parameters iteratively. In the present implementation, all RLC nodes are identical, and the coupling elements are controlled through discretized tuning elements with a finite number of levels. This establishes a fully in-situ optimization loop in which both the system response and its sensitivity are obtained directly from the physical platform.

Broadband engineering through parameter optimization is first explored and shown in Fig. 2B. The objective is defined to maximize the energy transfer over the entire time of measurement duration at node 3 when the system is excited from node 1. Starting from an initial configuration with modulated Gaussian pulse excitation and following adjoint measurement, the measured gradients, obtained from the interaction between the forward and adjoint responses, provide parameter updates over successive iterations, leading to parameter updates of the coupling elements. As optimization proceeds, the objective function increases monotonically, indicating consistent improvement of the system response. The resulting output spectrum (see low-right subfigure of Fig. 2B) exhibits a maximized energy transfer over broadband spectrum, demonstrating that the proposed framework can directly shape frequency-domain behavior through time-domain measurements.

Next, time-windowed energy control is demonstrated in Fig. 2C, by maximizing the difference in energy between two separate time windows $w_1: 0 \leq t \leq 0.65T$ and $w_2: 0.65T \leq t \leq T$, ($T$ is the total duration of the experiment) at selected nodes of the network. The network is excited with slowly modulated Gaussian pulse at node 1 and the objective is chosen such that node 3 stores the maximum (minimum) energy during $w_2$ ($w_1$) while node 4 accumulates the maximum (minimum) energy during $w_1$ ($w_2$). The measured gradients guide parameter updates that progressively reshape the temporal response. As a result, the objective improves over iterations, and the output waveform becomes increasingly confined to the desired time window, confirming that the framework can control the temporal structure of the system response.

Due to the discretized nature of the coupling elements in the current hardware, parameter updates are applied in finite steps, and the optimization primarily follows the sign of the measured

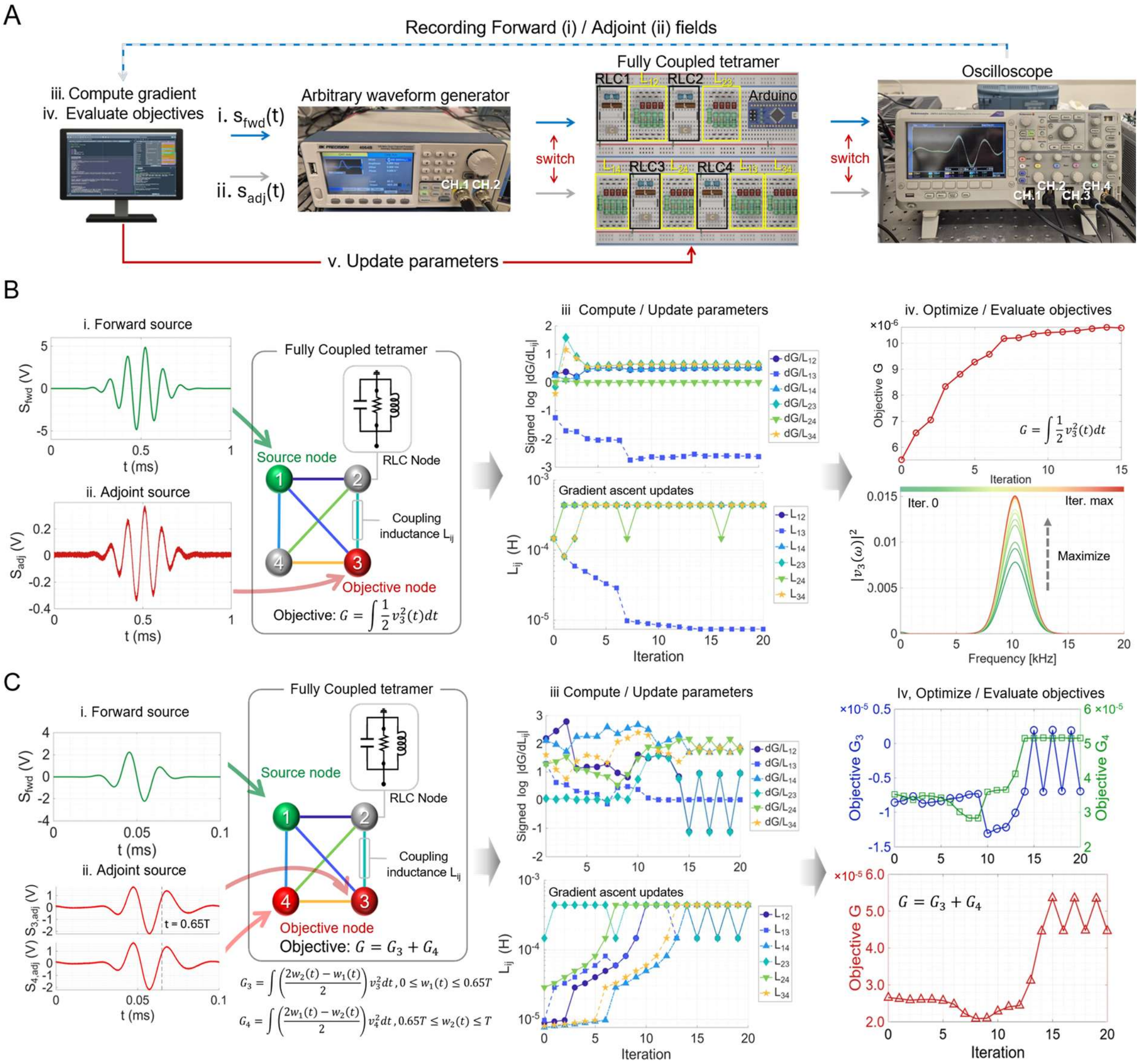

**Fig. 2 | Experimental in-situ adjoint optimization in a coupled RLC network.**

(A) Experimental implementation of parameter optimization. A programmable controller generates forward and adjoint excitations, controls switching, records the corresponding responses from a fully coupled RLC tetramer, computes gradients and objectives, and updates the tunable coupling inductances. Forward and adjoint fields are measured sequentially on the same physical network. (B) Broadband energy routing optimization. (i) A modulated Gaussian pulse is applied at node 1, while (ii) an adjoint excitation is applied at node 3. The coupling inductances are optimized to maximize the time-integrated energy at node 3, $G = \int \frac{1}{2} v_3^2(t)\, dt$. (iii) Experimentally measured adjoint gradients determine discrete updates of the coupling inductances. (iv) The objective increases over iterations, and the optimized spectrum shows enhanced broadband transfer to the target node. (C) Time-window optimization. (i) A shorter Gaussian excitation is applied at node 1, while (ii) an adjoint source is exciting nodes 3 and 4. The couplings are optimized to concentrate energy at nodes 3 and 4 within different temporal windows as described in the figure. The objective function implements a 'weight' factor of '2' to efficiently maximize the energy concentration at a designated node as shown in the objective function $G_3$ and $G_4$. (iii) The measured gradients guide the system toward coupling inductance configurations that reshape the transient responses into the desired time windows. (iv) Together, the broadband and time-window tasks demonstrate that in-situ adjoint gradients can optimize experimentally realized dynamics without reconstructing the full system model.

gradient rather than its magnitude. Despite this constraint, optimization consistently converges toward the desired responses in both broadband and time-window engineering tasks, indicating that the gradient direction alone is sufficient to guide the system toward optimal configurations in this implementation. In Extended Figure 1, we show the behavior of optimization with continuous tuning, which proves that the discretized step of inductance also provides the converged optima.

To validate the in-situ adjoint implementation, we perform numerical optimization on the same network model using the exact mathematical adjoint formulation in Extended Figure 2, where backward propagation in time can be implemented without the experimental constraints in the variation of the control parameters. In this case, the adjoint dynamics follow the ideal continuous formulation, providing a reference for comparison with the experimental implementation. We compare the optimization trajectories and final optimized responses obtained from experiment and numerical simulation. We observe close agreement between the two, with both approaches converging to nearly identical parameter configurations and achieving the same target responses in broadband and time-window engineering tasks. This agreement confirms that the proposed in-situ framework faithfully reproduces the behavior of the mathematical adjoint method, despite the constraints of physical implementation.

We next demonstrate in-situ source optimization, where the design variable is the time-dependent input waveform rather than the system parameters. In contrast to parameter optimization, source optimization requires a time-resolved sensitivity defined at each instant of time. This introduces a causality constraint because the objective is evaluated after the source excitation, preventing direct implementation of the mathematical adjoint source. The experimental implementation is illustrated in Fig. 3A. A forward experiment first generates the system response to an initial input waveform, and the resulting dynamics are evaluated through a time-dependent objective function. The objective-dependent adjoint source is then constructed directly from the measured forward response. To enforce causality, we introduce an additional experimental time reverse based on $(t = T - t)$, which is implemented by replaying the recorded signal in a causality-preserving form to generate the adjoint excitation over the full temporal duration. After the adjoint experiment, the measured adjoint response is reversed back to the original time coordinate, enabling reconstruction of the time-resolved sensitivity. We show that this time reversed adjoint source excitation (from the objective node) is equal to the time reversed adjoint field obtained from the standard adjoint source excitation in Supplementary Information Section 2.3.1[35]. The gradient with respect to the input waveform is then obtained from the reconstructed adjoint response through the $\Theta$ transformation, yielding $dG/ds(t)$ at every instant in time. This establishes a fully in-situ framework for waveform optimization, in which both forward and adjoint processes are realized and measured directly within the physical system.

We demonstrate this approach through time-dependent waveform shaping in both single-source and multi-source configurations using a chain-coupled tetramer system. In the single-source example shown in Fig. 3B, the system is excited at node 1 with a chopped Gaussian voltage waveform under 5 V, and the objective is to enhance and suppress, or enhance the difference of the squared voltage at node 3 within two prescribed temporal windows, $w_1(t)$ and $w_2(t)$. Starting from an initial waveform, the measured time-resolved gradients iteratively update the source excitation, progressively increasing the desired contribution while suppressing the response within the undesired window. As a result, both the input waveform is optimized toward the target temporal behavior. To ensure stable optimization during maximization, the source amplitude is constrained to a maximum of 5 V. We further extend the framework to multi-source waveform optimization with multiple objective nodes, as shown in Fig. 3C. Two independent inputs, consisting of triangular and rectangular pulses, are applied at nodes 1 and 3. The objective is to suppress the energy at node 2 within time window $w_1(t)$ and at node 4 within $w_2(t)$. The measured gradients drive simultaneous updates of both input waveforms, progressively reducing the target responses within the specified temporal windows. These results demonstrate that experimentally measured adjoint gradients can directly synthesize desired spatiotemporal input signals, extending in-situ adjoint optimization beyond parameter tuning to waveform engineering without reconstructing the underlying system model. To validate the in-situ adjoint implementation, we perform numerical optimization on the same network model using the exact mathematical adjoint formulation, see Extended Figure 3.

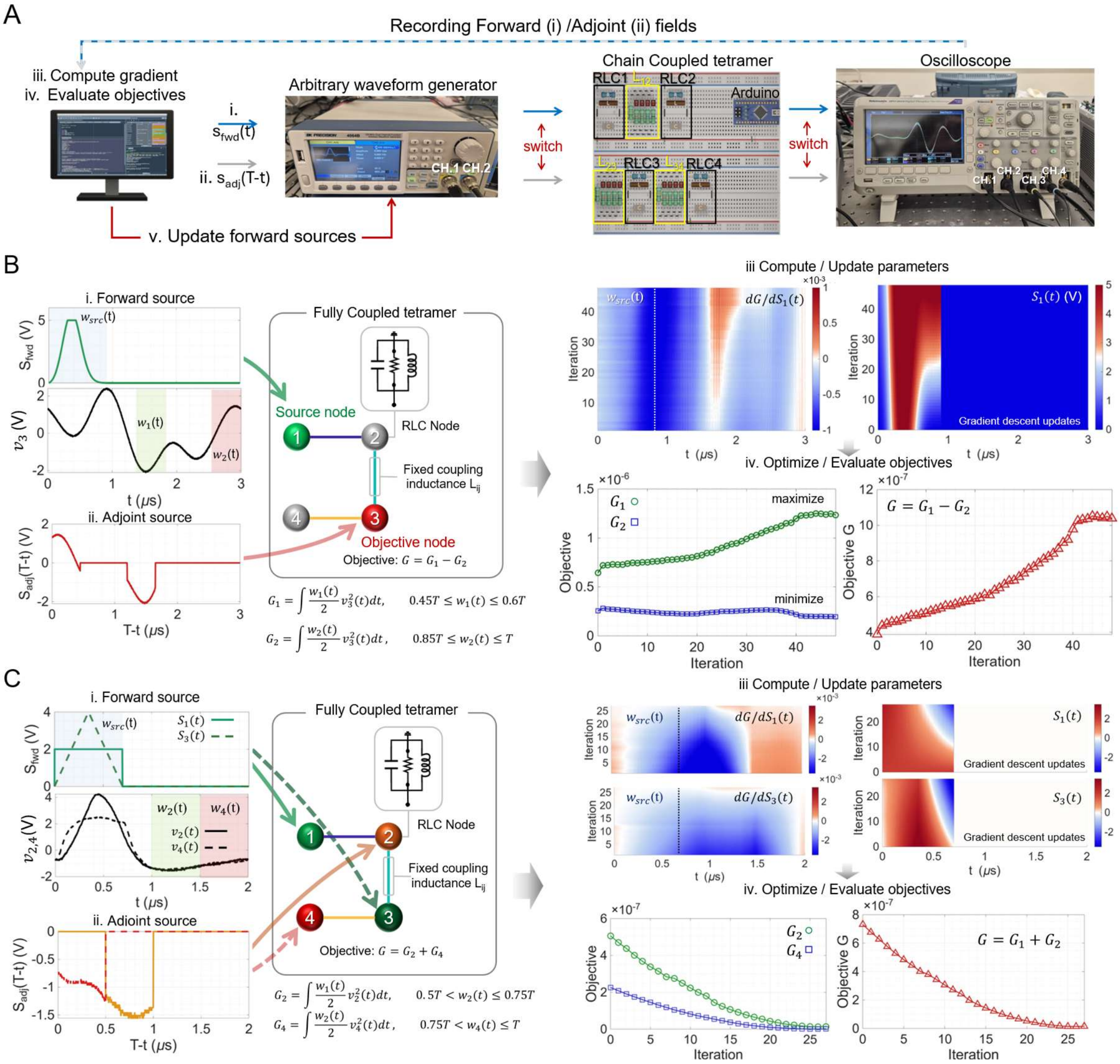

**Fig. 3 | Experimental in-situ source optimization using measured adjoint gradients.**
(A) Experimental implementation of source optimization. A programmable controller generates forward excitation, constructs the causality-preserving adjoint excitation from the recorded response, performs forward and adjoint measurements on the same coupled RLC network, computes time-resolved gradients, and updates the input waveform iteratively. The adjoint source is generated through experimental time remapping, enabling reconstruction of the source sensitivity directly from measured signals. A small apparent amplitude response at $t = 0$ arises from timing delay and synchronization jitter between the computer-controlled waveform generation and the measurement device (see the lower panels of Fig. 3B(i) and Fig. 3C(i)). This artifact is treated as experimental noise and is not included as a physical system response. (B) Single-source waveform optimization. (i) A chopped Gaussian excitation (upper figure of Figure 3B(i)) is applied at node 1, and the waveform is optimized to enhance and suppress the response of node 3 within two prescribed temporal windows, which are indicated in the initial measured voltage in bottom figure of Figure 3B(i)). (ii) The adjoint excitation is injected at node 3. (iii) Measured adjoint gradients drive iterative updates of the source waveform, resulting in progressive reshaping of both the excitation and the system response toward the desired temporal behavior. (iv) Objective function per window and total objective function $G$ versus iteration. (C) Multi-source waveform optimization. (i) Independent source waveforms are applied simultaneously at nodes 1 and 3 (upper figure of Fig. 3C(i)) and optimized to suppress responses at nodes 2 and 4 within designated temporal windows (bottom figure of Fig. 3C(i)). (ii) The adjoint excitations are injected at nodes 2 and 4. (iii) Time-resolved gradients obtained from measured adjoint responses update both input waveforms in parallel, producing selective suppression of the target responses. (iv) Objective function per window and total objective function $G$ versus iteration.

The validity of this scheme has also been tested successfully for a variety of other modalities like broadband minimization (Extended Figure 4).

**In-silico Implementation of Time-Domain Adjoint Optimization Protocol for Large Control Parameter System**

Following the experimental demonstrations of in-situ parameter optimization (Fig. 2) and source optimization (Fig. 3), we next explore the broader design space accessible to the proposed framework using a digital twin of a complex RLC network. While the experimental platform is constrained by the number of controllable sources, measurable nodes, and discretized tuning elements, the underlying adjoint formulation is not subject to these limitations. Figure 4 illustrates a representative wave dynamic optimization task that becomes accessible as the number of controllable degrees of freedom is increased, including broadband energy transfer, temporal localization, spatiotemporal routing, energy extinction, multiple sources, and multiple objective nodes. We first consider parameter optimization in networks with increasing structural complexity (Fig. 4A–C). Starting from a one-dimensional chain (Fig. 4A), optimization of the coupling parameters produces directed energy transport and temporal localization by reshaping the propagation pathways of the network. In this numerical configuration, the resistance at each node is uniform while all other parameters are randomly given. The optimized coupling distribution guides energy toward selected target nodes within desired temporal intervals, resulting in structured spatiotemporal energy flow which forms a "W" shape. Extending the same framework to a two-dimensional lattice (Fig. 4B) provides additional routes and enables stronger localization of energy in both space and time. The resulting optimized state exhibits concentration of energy at selected nodes while suppressing propagation elsewhere at specific time window. In a fully connected network (Fig. 4C), the optimization can simultaneously coordinate a large number of interacting pathways, enabling high-dimensional spatiotemporal control involving multiple target nodes and multiple temporal objectives. Together, these examples demonstrate that the proposed framework remains effective across network architectures ranging from sparse chains to densely connected systems and is not restricted to a particular topology.

Source optimization provides a complementary route to wave control in which the network structure remains fixed, and the excitation itself becomes the design variable. As shown in Fig. 4D, optimization of the source waveform enables selective suppression of energy at prescribed nodes and temporal windows without modifying the physical network. In this example, the optimized excitation progressively eliminates the response at the target node during the specified temporal interval, demonstrating spatiotemporal energy extinction through waveform engineering alone.

More generally, because both the source distribution and the objective function can be extended to multiple inputs and multiple target nodes, the framework naturally supports high-dimensional waveform synthesis and distributed energy control across complex networks. Together, these numerical examples show that the proposed adjoint framework is not limited to a specific topology, optimization variable, or objective type. By increasing the number of controllable sources, tunable parameters, and objective nodes, the same methodology can support broadband optimization, spatiotemporal localization, energy routing, energy extinction, and multi-objective waveform control in complex dynamical systems.

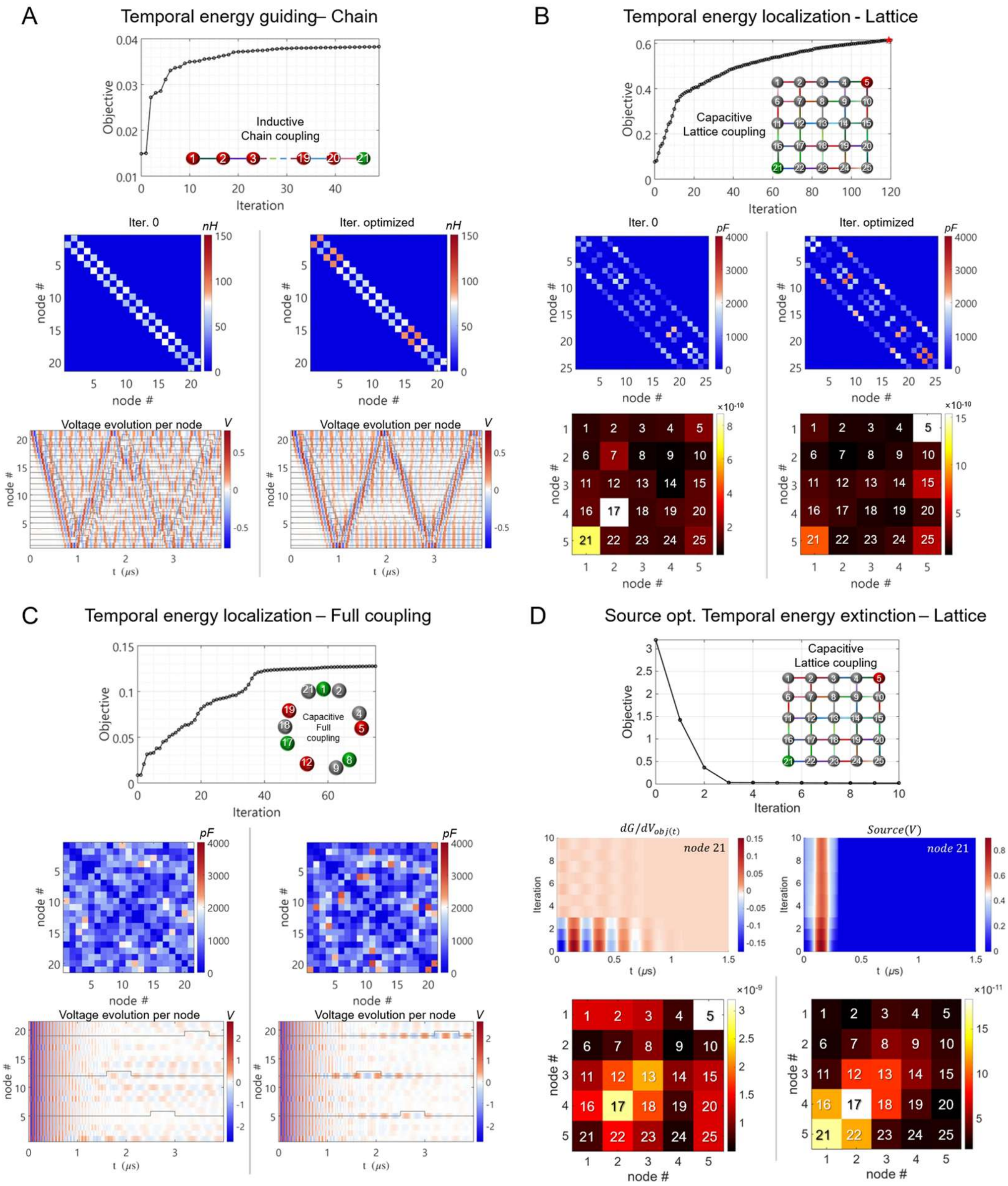

**Fig. 4 | Topology-agnostic wave control via system and source optimization**
For panels A-C, the top row shows the objective evolution, the middle row compares the coupling matrices before and after optimization, and the bottom row shows the corresponding voltage dynamics. Panel D instead shows the time resolved source sensitivity and source waveform evolution for source optimization. (A) System optimization in a one-dimensional inductive chain. A $40ns$ Gaussian pulse excites node 21 over a 4 μs duration, and the coupling coefficients are optimized to deliver energy to prescribed target nodes within selected temporal windows. The optimized couplings form directed pathways that guide energy along the chain. (B) System optimization in a two-dimensional capacitive lattice. With excitation at node 21 and node 5 as the target, optimization localizes energy within the $[1.5\mu s, 4\mu s]$ temporal window. The colormap shows the dissipated energy distribution across the lattice. (C) System optimization in a fully connected network. Excitations are applied at nodes 1, 8, and 17, while the coupling parameters are optimized to localize energy at nodes 5, 12, and 19 over distinct temporal windows, demonstrating spatiotemporal control without a predefined sparse topology. (D) Source optimization in a two-dimensional lattice with fixed coupling. The input waveform is optimized to suppress energy within the $[1\mu s, 1.5\mu s]$ temporal window at selected node 5, demonstrating spatiotemporal extinction through waveform control alone.

## Discussion and Conclusion

In this study, we developed a theoretical framework and presented a proof-of-concept experimental demonstration of an in-situ time-domain adjoint optimization for direct, measurement-driven control of dynamical physical systems. By realizing time-domain adjoint sensitivities on the same physical platform used for the forward evolution, the method enables gradient-based optimization without explicit reconstruction of the system model. We experimentally demonstrated parameter optimization for broadband and time-windowed response engineering as well as source optimization for temporal waveform shaping in complex multi-scattering networks of RLC resonators. Digital twin studies further showed that the formulation extends to larger and more highly connected networks, supporting multiple sources, objectives, and spatiotemporal energy routing tasks.

In contrast to an earlier frequency-domain in-situ adjoint optimization[13], the present work establishes a time-domain formulation for broadband and temporally resolved control. Our scheme can also be contrasted with existing time-reversal mirrors (TRM) protocols [15-17]. TRM records waveforms and re-emits them through a fixed scattering environment to achieve refocusing. Our approach instead realizes a transformed, objective-dependent adjoint propagation to obtain gradients for modifying the environment itself or optimizing the source excitation. Moreover, conventional TRMs rely on a time-reversal compatible propagation environment, while the present scheme operates even in the presence of losses.

A current limitation concerns operation in dynamically varying or noisy environments. Because the protocol relies on experimentally acquired forward and adjoint responses, the effects of temporal fluctuations and noise on optimization robustness remain to be systematically assessed. Addressing this limitation will be important for extending the approach to systems whose properties evolve during the optimization process.

Together, our experimental and numerical results establish a topology- and modality-independent route to physical optimization and provide a foundation for adaptive, real-time control of complex wave and network systems.

## Methods

### Circuit Design and Implementation

The circuit diagram of the fully coupled tetramer used in parameter optimization is shown in Extended Figure 5A. Each RLC unit consists of an inductor $L_i = 240\ \mu H$, a capacitor $C_i = 100\ nF$, and a resistor $R_i = 1\ k\Omega$, where $i \in \{1,2,3,4\}$. The RLC tanks are connected through an inductive series coupling implemented by a parallel combination of a base coupling inductor $L_{base}^{(ij)} = 440\mu H$ and four switchable coupling inductors with values $L_{swit}^{ij,1} = 10\ \mu H, L_{switc}^{ij,2} = 56\ \mu H, L_{switch}^{ij,3} = 100\ \mu H$, and $L_{switch}^{ij,4} = 220\mu H$. Together, the four switching states and the base inductor provide $2^4 = 16$ distinct effective coupling inductance values for each pair of RLC units. The circuit diagram of the chain-coupled tetramer used in source optimization is shown in Extended Figure 5B. Each RLC unit consists of an inductor $L_i = 120\mu H$, a capacitor $C_i = 1000\ pF$, and a resistor $R_i = 1\ k\Omega$. The same inductive ladder architecture is used for each adjacent coupling link, but the smallest switchable inductance is changed from $10\mu H$ to $20\ \mu H$.

To ensure equivalent boundary conditions under the forward and adjoint excitation, a pair of switches interchange the source and termination ports of the network. During the forward measurement, switch $S_F$ connects the arbitrary waveform generator (AWG) to RLC unit 1 through its $50\Omega$ output impedance, while the target RLC unit $i$ is connected through switch $S_A$ to an equivalent $50\Omega$ termination implemented using two parallel $R_T = 100\Omega$ resistors connected to the ground. During the adjoint measurement, the switch configuration is exchanged. Switch $S_A$ connects the AWG to the target RLC unit, where the adjoint excitation is applied, while switch $S_F$ connects RLC unit 1 to the equivalent $50\Omega$ termination. This configuration preserves identical external resistive loading conditions during the forward and adjoint measurements while avoiding additional reactive phase shifts. The component models used in the experiment are provided in the Extended Table 1, while more information on the circuit is given in Supplementary Information Section 4[35].

The switch is powered by a $\pm 5\ V$ power supply (Rigol DP932E). The forward and adjoint source waveforms were generated by a B&K Precision 4064B AWG, operated at an instrument maximum sampling rate of $75\ MSa/s$ to transmit high-fidelity forward and adjoint source excitations. The time-domain data recorded at each node were acquired by a Tektronix DPO2014B oscilloscope with sampling rate ranging from $125\ MSa/s$ to $1\ GSa/s$. The bandwidth of the oscilloscope was set to $20\ MHz$ to filter out high-frequency noise sources from the recorded signal. The device specifications are provided in Extended Table 2.

**General Principles of In-situ Time Adjoint Optimization**

The system equations can be written in implicit form as $F\left(\psi, \dot{\psi}, p, s(t), t\right) = 0$ with initial condition $\psi(0) = \psi_0$. We introduce a Lagrange multiplier $\phi(t)$, denoted as an adjoint variable, and define an augmented function as,

$$I\left(\psi, p, s(t)\right) = G\left(\psi, p, s(t)\right) - \int_0^T \phi^T(t)\, F\left(\psi, \dot{\psi}, p, s(t), t\right) dt,$$

Then the sensitivity of the objective function can be written as

$$\frac{dG\left(\psi, p, s(t)\right)}{dp} = \frac{dI\left(\psi, p, s(t)\right)}{dp} = \int_0^T \left(\frac{\partial g}{\partial p} + \frac{\partial g}{\partial \psi}\frac{\partial \psi}{\partial p}\right) dt - \int_0^T \phi^T \left(\frac{\partial F}{\partial p} + \frac{\partial F}{\partial \psi}\frac{\partial \psi}{\partial p} + \frac{\partial F}{\partial \dot{\psi}}\frac{\partial \dot{\psi}}{\partial p}\right) dt.$$

Integrating by parts the term containing $\partial\dot{\psi}/\partial p$ gives

$$\frac{dG\left(\psi, p, s(t)\right)}{dp} = \int_0^T \left(\frac{\partial g}{\partial p} - \phi^T \frac{\partial F}{\partial p}\right) dt - \int_0^T \left[-\frac{\partial g}{\partial \psi} + \phi^T \frac{\partial F}{\partial \psi} - \left(\phi^T \frac{\partial F}{\partial \dot{\psi}}\right)^{\cdot}\right] \frac{\partial \psi}{\partial p} dt - \left(\phi^T \frac{\partial F}{\partial \dot{\psi}}\frac{\partial \psi}{\partial p}\right)\Bigg|_0^T,$$

which allows us to derive the adjoint equation Eq. (3) in the main text by setting the term inside the square bracket to zero. For a pure running objective in time, $\phi(T) = 0$. Similarly, for fixed initial conditions that are independent of $p$, we get $\partial\psi_0/\partial p = 0$. These conditions are satisfied by our experiments; thus, resulting to a vanishing boundary term. Together with the fact that $\frac{\partial g}{\partial p} = 0$, the above equation leads to the expression Eq. (2) for the sensitivity.
Next, we comment on the required time reparameterization and transformation map that are required for the physical implementation of the adjoint scheme. Conventional time adjoint optimization is naturally expressed as a terminal value problem, because the adjoint state is constrained by information defined at the end of the observation window. Although this formulation is commonly interpreted as requiring backward temporal evolution, the apparent reversal is a consequence of mathematical parametrization rather than a requirement for physical propagation backward in time. We therefore introduce a reparametrized adjoint time coordinate,

$$\tau = T - t_{math},$$

which maps the terminal point of the mathematical adjoint problem onto the initial point of a new forward evolving coordinate. In this representation, the adjoint information generated by the objective can be launched as an experimentally prescribed initial value problem and allowed to evolve causally from 0 to $T$ on the laboratory clock. Thus, the terminal condition is not removed; rather, its role is transferred to the initial state of the physical adjoint experiment, which can be easily removed by defining objective function (See Supplementary Information section 1.2.3) [35]. The reparametrized dynamics, however, remain governed by the transposed system operator which in general differs from the operator realized by the physical system. To establish a correspondence between these descriptions, we introduce a linear state transformation $\Theta$ satisfying

$$H = \Theta H^{\mathrm{T}} \Theta^{-1}; \Theta = \begin{bmatrix} 0 & \mathbb{I} \\ \mathbb{I} & -\boldsymbol{G} \end{bmatrix},$$

Where $\boldsymbol{G} = (1/R)\mathbb{I}$ is the conductance matrix and $\mathbb{I}$ is the identity. This relation maps the mathematical adjoint dynamics governed by $H^T$ onto physically accessible variables evolving under the original operator $H$. The $\Theta$ transformation therefore provides a physical representation of the adjoint state rather than modifying the underlying dynamics of the system itself. Together, the time reparametrization and $\Theta$ transformation convert the conventional terminal value adjoint formulation into a causal initial value evolution that can be executed on the same physical platform used for the forward experiment. Both measurements are consequently performed along a forward laboratory time axis, and no physical backward propagation or post acquisition reversal of the measured fields is required. This correspondence allows the physical system to perform the dynamical operation required for adjoint sensitivity evaluation directly, forming the basis of the in-situ time adjoint optimization framework. The detailed explanation is provided in Supplementary Information Section 1 [35].

**Circuit Modeling using Kirchhoff's Law**

The forward problem of inductively coupled RLC network can be developed with Kirchhoff's current law at each node. If we apply this at node $i$, which is coupled to all other nodes $j$ with inductance $L_{m,ij}$, the current source $I_i(t)$ that injected at node $i$ should satisfy the relation

$$C_i\dot{v}_i + G_i v_i + \frac{1}{L_i}\int_0^T v_i \cdot dt + \sum_{i\neq j}\frac{1}{L_{m,ij}}\int_0^T (v_j - v_i)\cdot dt = I_i(t); \quad v_i = L_{tot}\dot{i}_{L,i}$$

where $v_i$ is the voltage across the $i-$th node, $i_{L,i}$ is the inductive current going through the inductor of this node. and $L_{tot}$ is the total inductance matrix of coupled system. This equation can be further expressed in matrix form for all nodes' voltage $\boldsymbol{v}$ and inductive current $\boldsymbol{i}_L$ providing the following forward state-space equation

$$\mathcal{F}: \dot{\boldsymbol{\psi}}(t) + H(\boldsymbol{p})\boldsymbol{\psi}(t) - \boldsymbol{C}^{-1}\boldsymbol{I}_i(t) = 0; \quad H(\boldsymbol{p}) = \begin{bmatrix} \boldsymbol{C}^{-1}\boldsymbol{G} & \boldsymbol{C}^{-1} \\ -\boldsymbol{L}_{tot,inv} & \boldsymbol{0} \end{bmatrix}$$

where $\psi(t) = [v_1, v_2, \cdots v_{n-1}, v_n, i_{L,1}, i_{L,2} \cdots i_{L,n-1}, i_{,L,n}]^T$ is the state-space vector, and $I_i(t) = [I_1(t), I_2(t), \cdots I_n(t), 0,0, \cdots 0]^T$ are the injected currents in the $i-$th node of the network. The submatrices that compose the Jacobian $H(\boldsymbol{p})$ take the form

$$(L_{tot,inv})_{ij} = \begin{cases} \left(\frac{1}{L_i} - \sum_{k=1,k\neq i}^{n} \frac{1}{L_{m,ik}}\right), & i = j \\ \frac{1}{L_{ij}}, & i \neq j \end{cases}, \qquad G_{ij} = \frac{1}{R_i}\delta_{ij}, \qquad C_{ij} = C_i\delta_{ij}$$

where $\delta_{ij}$ is the Kronecker-delta. In our setting, the resistance is assumed to be identical for all nodes, so that the conductance matrix is proportional to the identity matrix, $\boldsymbol{G} = (1/R)\mathbb{I}$. A detailed derivation and extension to the case of capacitively coupled RLC networks is presented in Supplementary Information Section 2 [35].

**Optimization Loop**

The full optimization procedure was implemented experimentally by alternating forward and adjoint measurements and reconstructing the required gradients directly from the measured temporal responses. For parameter optimization, the coupling strengths of the fully connected RLC tetramer were initialized and subsequently updated iteratively while the forward excitation waveform was held fixed. At each iteration, the prescribed waveform was injected into node 1 and the nodal voltages of the tetramer were recorded, with the target node terminated under the prescribed load condition. The measured response at the target node was then used to construct the adjoint excitation. Before the adjoint measurement, the source and target connections were interchanged using switches so that the forward and adjoint experiments experienced equivalent terminal loading conditions, while all coupling parameters were kept unchanged. The adjoint excitation was then applied through the target node, and the resulting adjoint nodal voltages were recorded and combined with the forward measurements to evaluate the sensitivity of the objective function with respect to the coupling parameters. The couplings were updated according to the calculated gradient until the objective saturated or the experimentally accessible parameter bounds were reached.

Source optimization followed the same forward and adjoint measurement protocol, but with the coupling configuration fixed throughout the optimization. In this case, the experimentally applied source waveform constituted the optimization variable, and the measured target response was used to construct the adjoint excitation. The resulting adjoint measurements were used to evaluate the functional gradient with respect to the source waveform, which was iteratively updated subject to the accessible voltage bounds until convergence. The detailed description of optimization loop is provided in Supplementary Information Section 4 [35].

**Objective functions for Spatiotemporal Optimization in Fig. 4**

The objective function was defined to temporally concentrate the potential energy of voltage squared at a prescribed target node while suppressing responses at all other nodes. For each temporal window $w_r(t)$, the duration of the window was defined as

$$|w_r| = \int w_r(t)dt,$$

The target response within each window was quantified by the window normalized mean squared voltage,

$$E_{obj}(l) = \frac{1}{|w_r|}\int w_r(l,t)v_{out}^2(l,t)dt\,, \quad l = 1, \cdots, n$$

To suppress undesired responses throughout the remainder of the network, the corresponding mean squared voltage of all non-target nodes was defined as

$$E_{all\ othe}\quad(l) = \frac{1}{|w_r|}\int \frac{w(t)}{n-1} v_{out}^2(l,t)dt,$$

The total objective function was obtained by averaging the target enhancement and non-target suppression over all $R$ number of temporal windows,

$$G = \sum_{r=1}^{R}\left(E_{obj}^{\mathrm{r}} - \lambda E_{all\ others}^{r}\right),$$

where $\lambda$ is the penalty weight for the suppression of all other nodes. Maximizing or minimizing $G$ therefore enhances or reduces the voltage response of the designated target node within each assigned temporal window while simultaneously suppressing or increasing the response at the remaining nodes.

## FUNDING

This work was partially supported by MURI under award number 23 000005209, AI Guided Self Organization: Tailoring Disorder to Shape Complex Nonlinear Dynamics, and by the Simons Foundation under grant SFI MPS EWP 00008530 08.

## ACKNOWLEDGEMENT

The authors thank Prof. Steven G. Johnson and Prof. Z. Lin for insightful discussions and feedback.

## REFERENCES

1. Miller, D. A. B. Self-configuring universal linear optical component. Photon. Res. 1, 1–15 (2013).
2. Sone, K., Ashida, Y. & Sagawa, T. Topological synchronization of coupled nonlinear oscillators. Phys. Rev. Res. 4, 023211 (2022).
3. Hughes, T. W., Minkov, M., Williamson, I. A. D. & Fan, S. Training of photonic neural networks through in situ backpropagation and gradient measurement. Optica 5, 864–871 (2018).
4. Pai, S. et al. Experimentally realized in situ backpropagation for deep learning in photonic neural networks. Science 380, 398–404 (2023).

5. Wright, L. G. et al. Deep physical neural networks trained with backpropagation. Nature 601, 549–555 (2022).
6. Momeni, A. et al. Training of physical neural networks. Nature 645, 53–61 (2025).
7. Stern. M, et al., Physical learning beyond the quasistatic limit, Physical Review Research 4, L022037, 2022.
8. Lions, J. L. Optimal Control of Systems Governed by PDEs (1971).
9. Errico, R. M. What is an adjoint model? Bull. Am. Meteorol. Soc. 78, 2577–2591 (1997).
10. Cao, Y. et al. Adjoint sensitivity analysis for differential algebraic equations: algorithms and software. J. Comput. Appl. Math. 149, 171–191 (2002).
11. Cao, Y. et al. Adjoint sensitivity analysis for differential algebraic equations: the adjoint DAE system and its numerical solution. SIAM J. Sci. Comput. 24, 1076–1089 (2003).
12. Li, S. & Mao, X. Training all-mechanical neural networks for task learning through in situ backpropagation. Nat. Commun. 15, 10528 (2024).
13. Guillamon, J. et al. In-situ physical adjoint computing in multiple-scattering electromagnetic environments for wave control. Nat. Commun. 16, 11466 (2025).
14. Li, J. et al. Circuit theory of time domain adjoint sensitivity. IEEE Trans. Comput. Aided Des. Integr. Circuits Syst. 42, 2303–2316 (2023).
15. Fink, M. Time reversal of ultrasonic fields. I. Basic principles. IEEE Trans. Ultrason. Ferroelectr. Freq. Control 39, 555–566 (1992).
16. Fink, M. Time reversal mirrors. In Acoustical Imaging Vol. 21 (ed. Jones, J. P.) 1–15 (Springer, 1995).
17. Fink, M. Time reversed acoustics. Phys. Today 50, 34–40 (1997)
18. Molesky, S. et al. Inverse design in nanophotonics. Nat. Photonics 12, 659–670 (2018).
19. Piggott, A. Y. et al. Fabrication-constrained nanophotonic inverse design. Sci. Rep. 7, 1786 (2017).
20. Liu, D. et al. Training Deep Neural Networks for the Inverse Design of Nanophotonic Structures, ACS Photonics 5, 1365–1369, (2018).
21. Shahriari, B., Swersky, K., Wang, Z., Adams, R. P. & de Freitas, N. Taking the human out of the loop: A review of Bayesian optimization. Proceedings of the IEEE 104, 148–175 (2016).
22. Wiecha, P. R., Arbouet, A., Girard, C. & Muskens, O. L. Deep learning in nano-photonics: inverse design and beyond. Photonics Research 9, B182–B200 (2021).
23. Nakajima, M. et al. Physical deep learning with biologically inspired training method. Nat. Commun. 13, 7847 (2022).
24. Sajedian, I., Badloe, T. & Rho, J. Finding the best design parameters for plasmonic nanostructures using genetic algorithms. Optics Express 27, 15141–15153 (2019).
25. Momeni, A. et al. Backpropagation-free training of deep physical neural networks. Science 382, 1297–1303 (2023).
26. Rotter, S., Gigan, S., Light fields in complex media. Rev. Mod. Phys. 89, 015005 (2017).
27. Mosk, A. P. et al. Controlling waves in space and time for imaging and focusing in complex media. Nat. Photonics 6, 283–292 (2012).
28. Fleury, R., Sounas, D. & Alù, A. Subwavelength ultrasonic circulator. Science 343, 516–519 (2014).
29. Vellekoop, I. M., Mosk, A. P., Focusing coherent light through opaque strongly scattering media, Optics Letters 32, 2309-11, 2007.
30. Popoff, S. M., Lerosey, G., Carminati, R., Fink, M., Boccara, A. C., and Gigan S., Measuring the Transmission Matrix in Optics: An Approach to the Study and Control of Light Propagation in Disordered Media, Phys. Rev. Lett., 104, 100601, 2010.
31. Rumelhart, D. E., Hinton, G. E. & Williams, R. J. Learning representations by back propagating errors. Nature 323, 533–536 (1986).
32. Bakr, M. H., Ahmed, O. S., El Sherif, M. H. & Nomura, T. Time domain adjoint sensitivity analysis of electromagnetic problems with nonlinear media. Optics Express 22, 10831–10845 (2014).
33. Tromp, J., Tape, C. & Liu, Q. Seismic tomography, adjoint methods, time reversal and banana doughnut kernels. Geophys. J. Int. 160, 195–216 (2005).
34. López-Pastor, V. & Marquardt, F. Self-Learning Machines Based on Hamiltonian Echo Backpropagation. Phys. Rev. X 13, 031020 (2023).
35. Kwon, H. et al. Supplementary Information.
36. Alexander, C. K., Sadiku, M. N. O. *Fundamentals of Electric Circuits*. (McGraw-Hill Education, 2021).

**Extended Figure 1.**

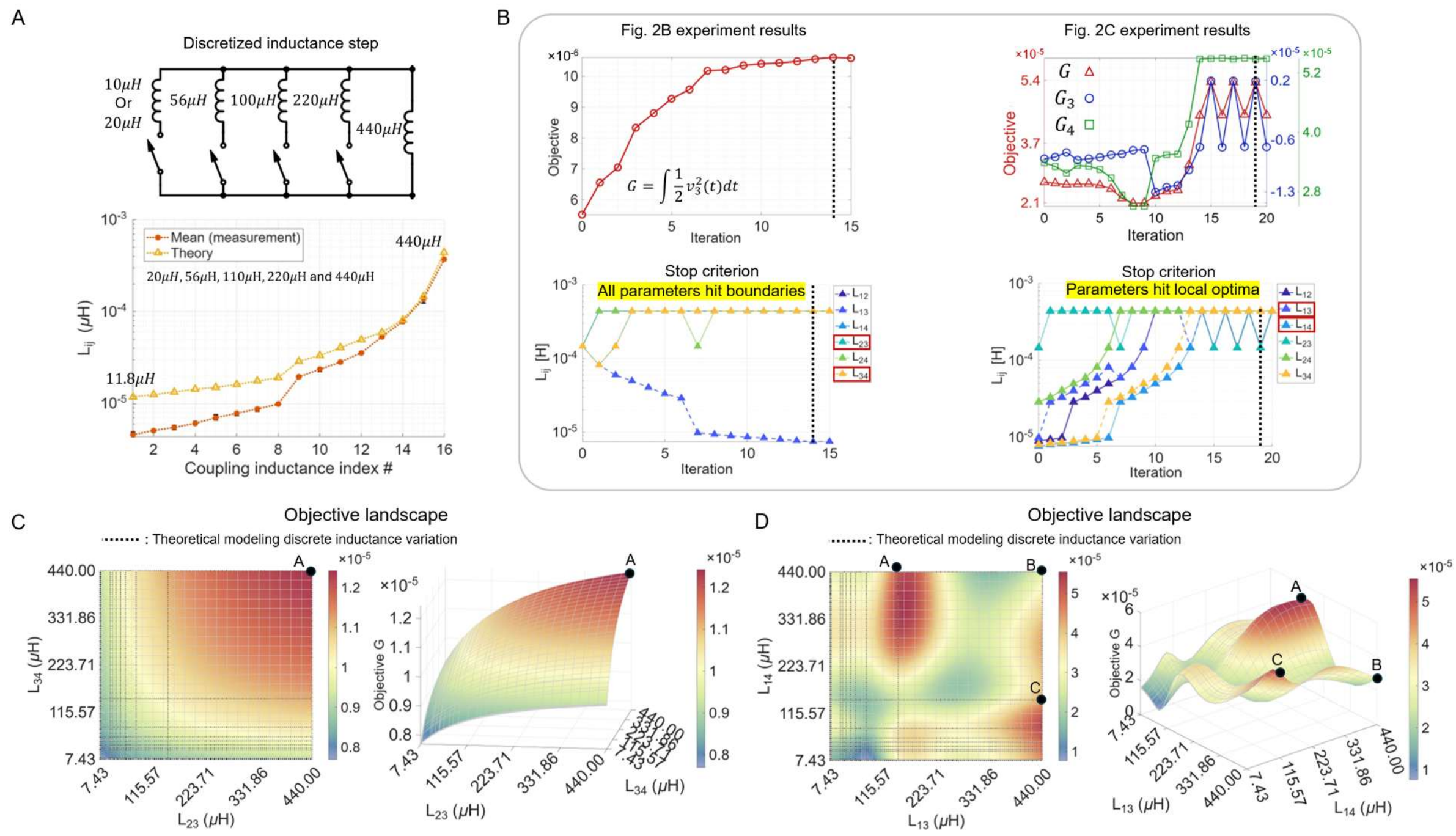

**Extended Figure 1 | Convergence behavior and objective landscape of discrete adjoint optimization.** (A) Experimental realization of discretized coupling inductances. The tunable inductance is implemented using a switched inductor network composed of 10 (20), 56, 100, and $220\mu H$ elements connected in parallel with a fixed $440\mu H$ inductor. The resulting 16 discrete inductance states span from 7.43 (11.8) to $440\mu H$. Measured inductance values (red line) are compared with theoretical modeling of discrete inductance variation (yellow line), demonstrating accurate realization of the discrete parameter space used during optimization. (B) Optimization trajectories for the two experimental demonstrations presented in Fig. 2B and Fig. 2C in the main text. The upper panels show the evolution of the objective function over successive optimization iterations, while the lower panels show the corresponding discrete parameter updates. In the Fig. 2B experiment, the optimization converges when all tunable parameters reach the boundary of the accessible parameter space. In the Fig. 2C experiment, the optimization terminates when no neighboring discrete state yields a higher objective value, indicating convergence to a local optimum. (C) Objective landscape associated with the experiment of Fig. 2B. Here, only $L_{34}$and $L_{23}$ are changed while all other coupling inductances are fixed at optimal value. Left panel: Simulation results of objective landscape with continuous parameter updates. The black dashed lines are associated with modeling of discrete step optimization that has been applied to experiment. Right panel: same data of upper left panel shown as 3D objective landscape. The experimentally obtained solution (point A) lies at the boundary of the parameter space and coincides with the global maximum of the scanned region, explaining the observed convergence behavior. (D) Objective landscape associated with the experiment of Fig. 2C. Here, only $L_{14}$ and $L_{13}$ are changed while all other coupling inductances are fixed at optimal value. Left panel: Simulation results of objective landscape with continuous parameter updates. The black dashed lines are associated with modeling of discrete step optimization that has been applied to experiment. Right panel: same data of upper left panel shown as 3D objective landscape. Unlike the monotonic landscape shown in (C), the objective function exhibits multiple extrema and non-convex features. The continuous parameter updates convex is reduced to discrete parameter updates landscape, giving local optimum A (black filled circle) within parameter space. Although gradient at A requires parameters towards increasing values B or C (black filled circles) in the discrete parameter space, the gradient at B or C rewinds the parameter values back to the local optimum A. Together, these results justify the discrete update strategy, establish the convergence criterion used in the experiments, and reveal how the topology of the objective landscape determines whether optimization terminates at a boundary optimum or a local optimum.

**Extended Figure 2.**

A. Digital twin results of Fig. 2B

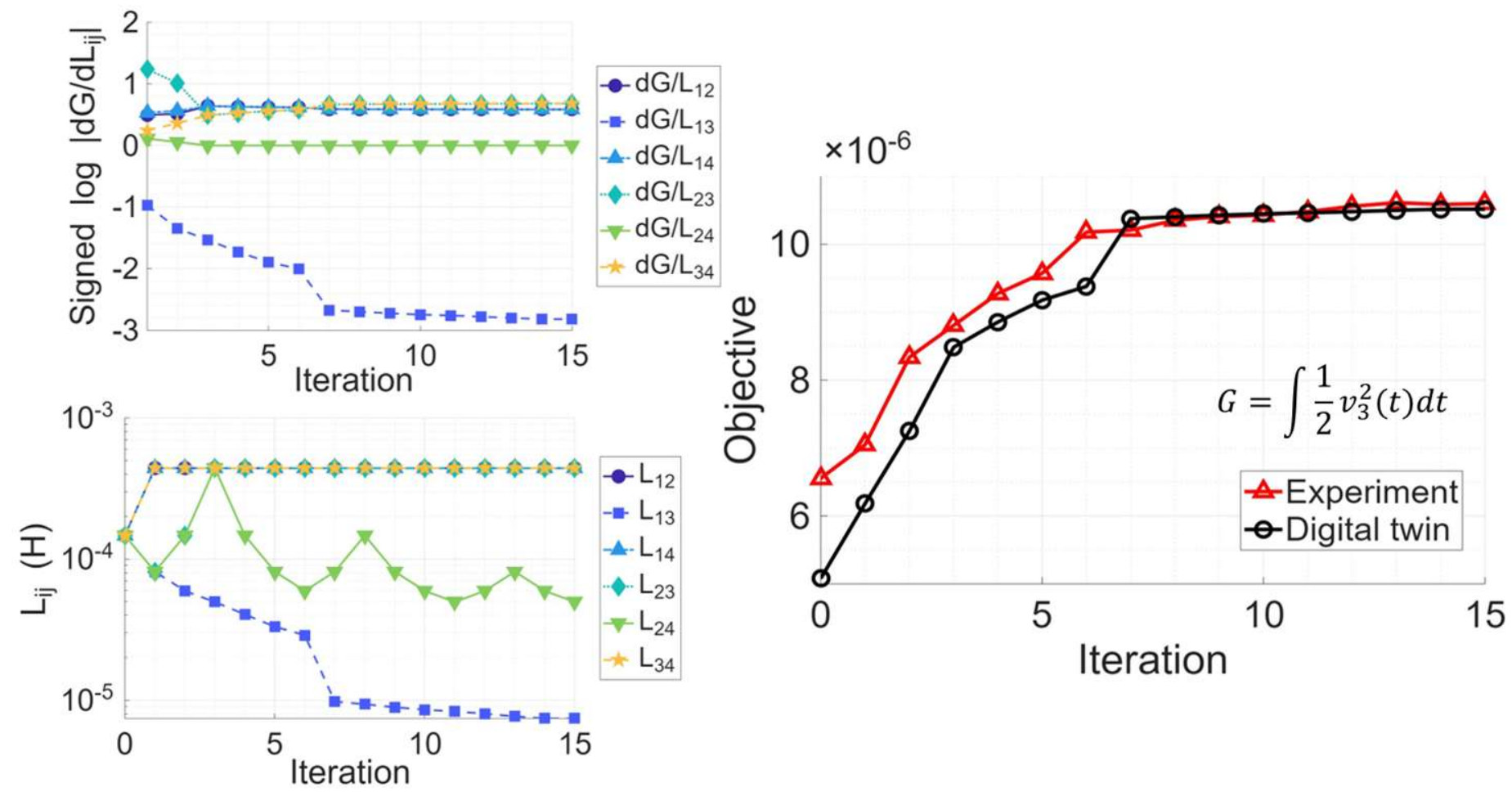

B. Digital twin results of Fig. 2C

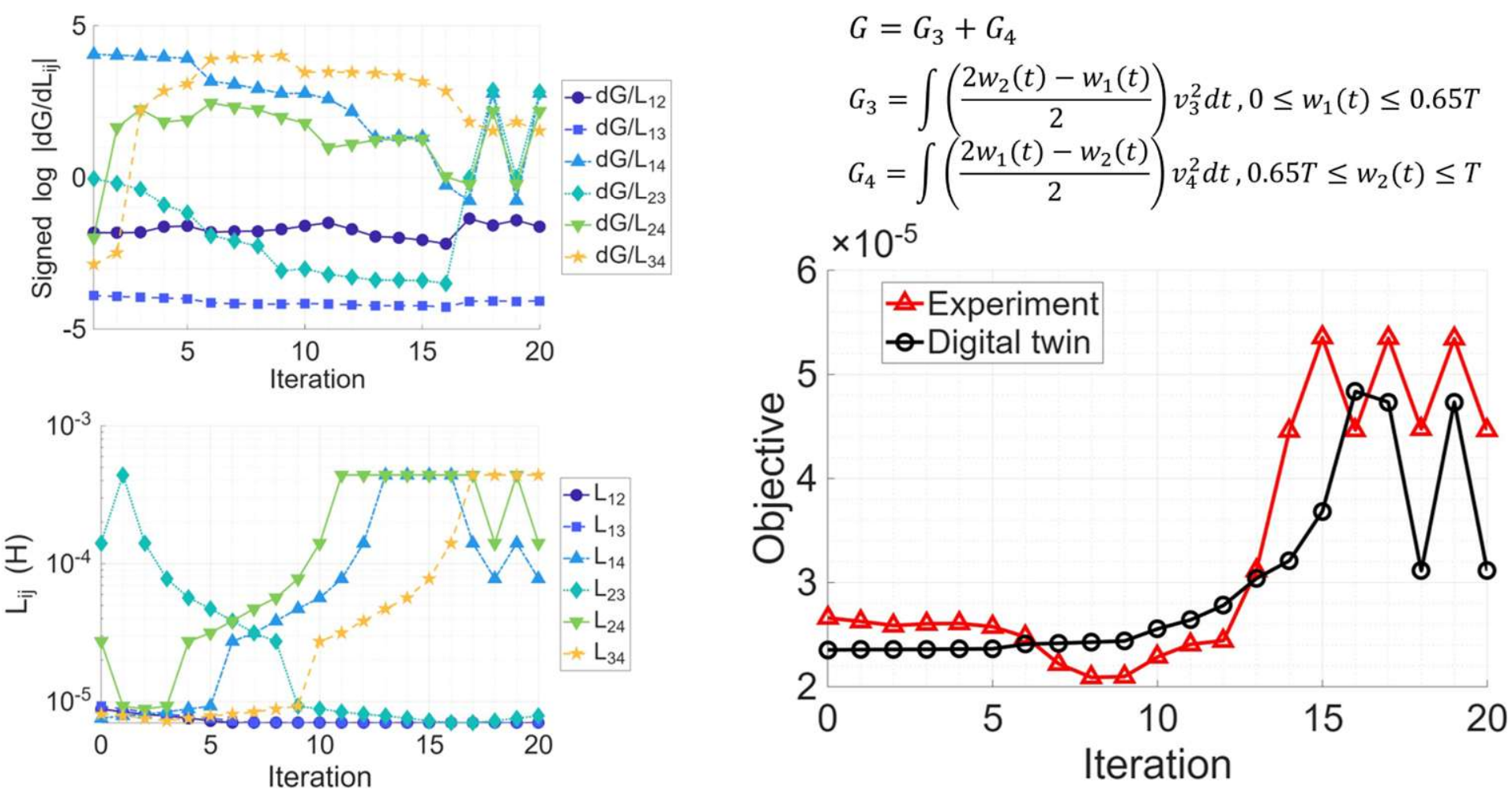

**Extended Figure 2 | Digital twin results of parameter optimization**

(A) Digital twin results corresponding to Fig. 2B (broadband maximization through coupling-inductance optimization). The left upper panel shows the evolution of the parameter sensitivities, and the left bottom panel provides coupling inductance updates over the optimization iterations. The digital twin accurately reproduces the experimentally observed sensitivity trends and inductance-update trajectories. The right panel shows the objective-function evolution, further confirming that the digital twin (black line-cycles) captures the convergence behavior observed in the experiment (red line-triangles). (B) Digital twin results corresponding to Fig. 2C (temporal energy routing). The left upper panel provides evolution of sensitivities, while the left bottom panel shows coupling-inductance updates. The right panel shows the objective per iteration as they are evaluated from the digital twin (black line-cycles) and the experiment (red line-triangles), which exhibits similar convergence trends throughout the optimization. The oscillating behavior of objective function in convergency is due to the fact that the optimization parameters hit the local optimum (See caption of Extended Figure 1D).

**Extended Figure 3.**

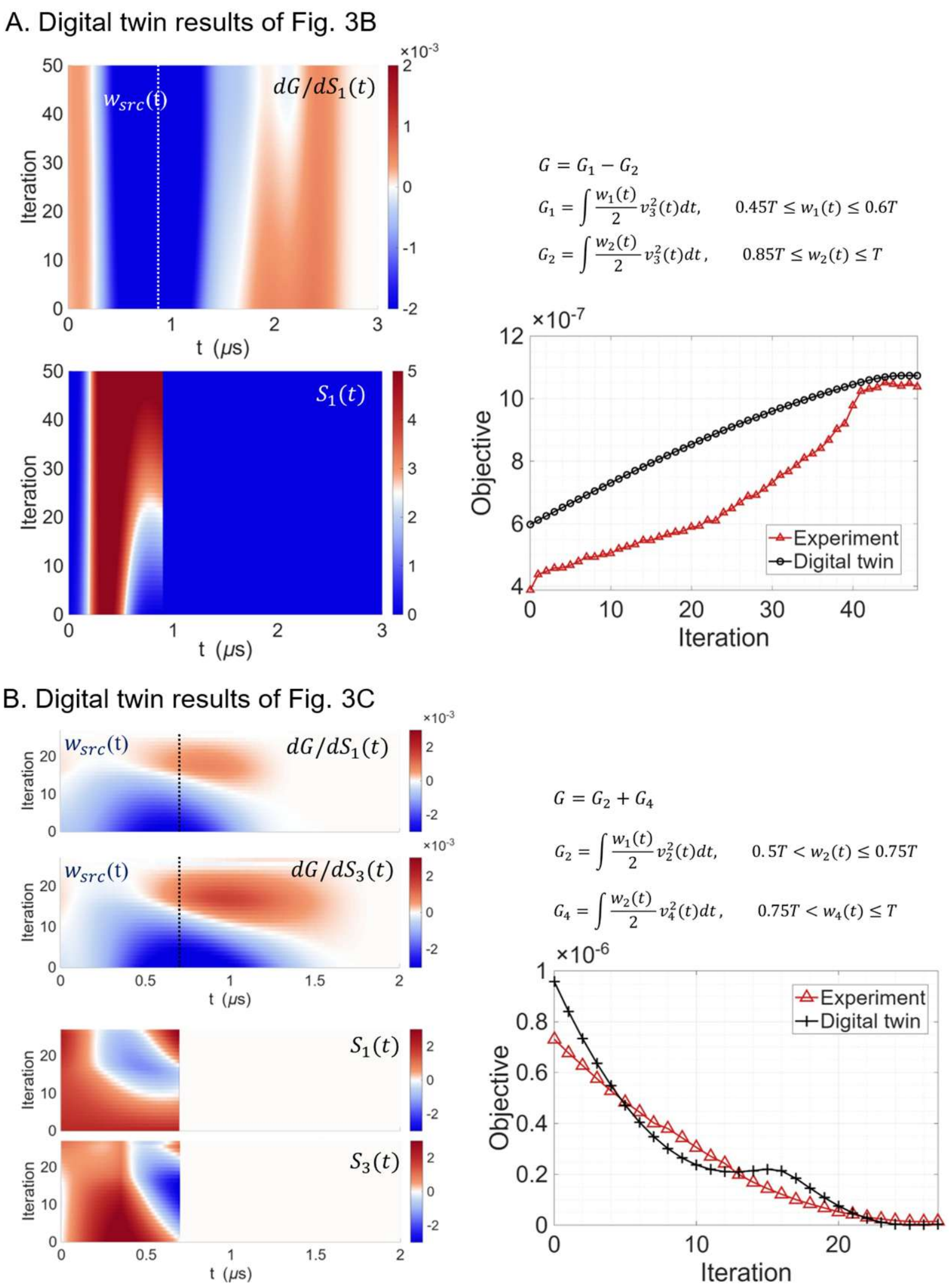

**Extended Figure 3 | Digital twin results of source optimization**

(A) Digital twin results corresponding to Fig. 3B (single-source waveform optimization). The left upper panel shows the evolution of the source sensitivity $dG/dS_1(t)$ over the optimization iterations. The time window $w_{src}(t)$ during which the source is applied is indicated with white dashed line. The bottom left panel shows the temporal progression of the source profile $S_1(t)$ which adapts according to the measured sensitivities, leading to enhanced energy delivery within the target time window $w_1(t)$: $0.45T \leq w_1(t) \leq 0.6T$ and suppressed energy delivery within $w_2(t)$: $0.85T \leq w_2(t) \leq T$. Also, the maximum voltage of the source is limited to 5V. The right panel shows the objective function trajectory, as it is evaluated from the digital twin, closely follows the experimental trend, confirming that the digital twin accurately reproduces the source-optimization dynamics and convergence behavior observed in the experiment. (B) Digital twin results corresponding to Fig. 3C (multi-source temporal energy routing). The left upper panel shows the sensitivities for the independently optimized source channels. The time window $w_{src}(t)$ during which the source is applied is indicated with black dashed lines. The left bottom panel provides the optimized source profiles which evolve to cooperatively redistribute energy in and suppress the response within the penalty window time $w_1(t)$: $0.5T \leq w_1(t) \leq 0.75T$ and $w_2(t)$: $0.75T \leq w_2(t) \leq T$. The right panel provides the evolution of the objective function obtained from the digital twin. The latter remains in close agreement with the experimental results throughout the optimization process, demonstrating that the digital twin accurately captures both the source-update dynamics and the resulting temporal energy-routing behavior.

**Extended Figure 4.**

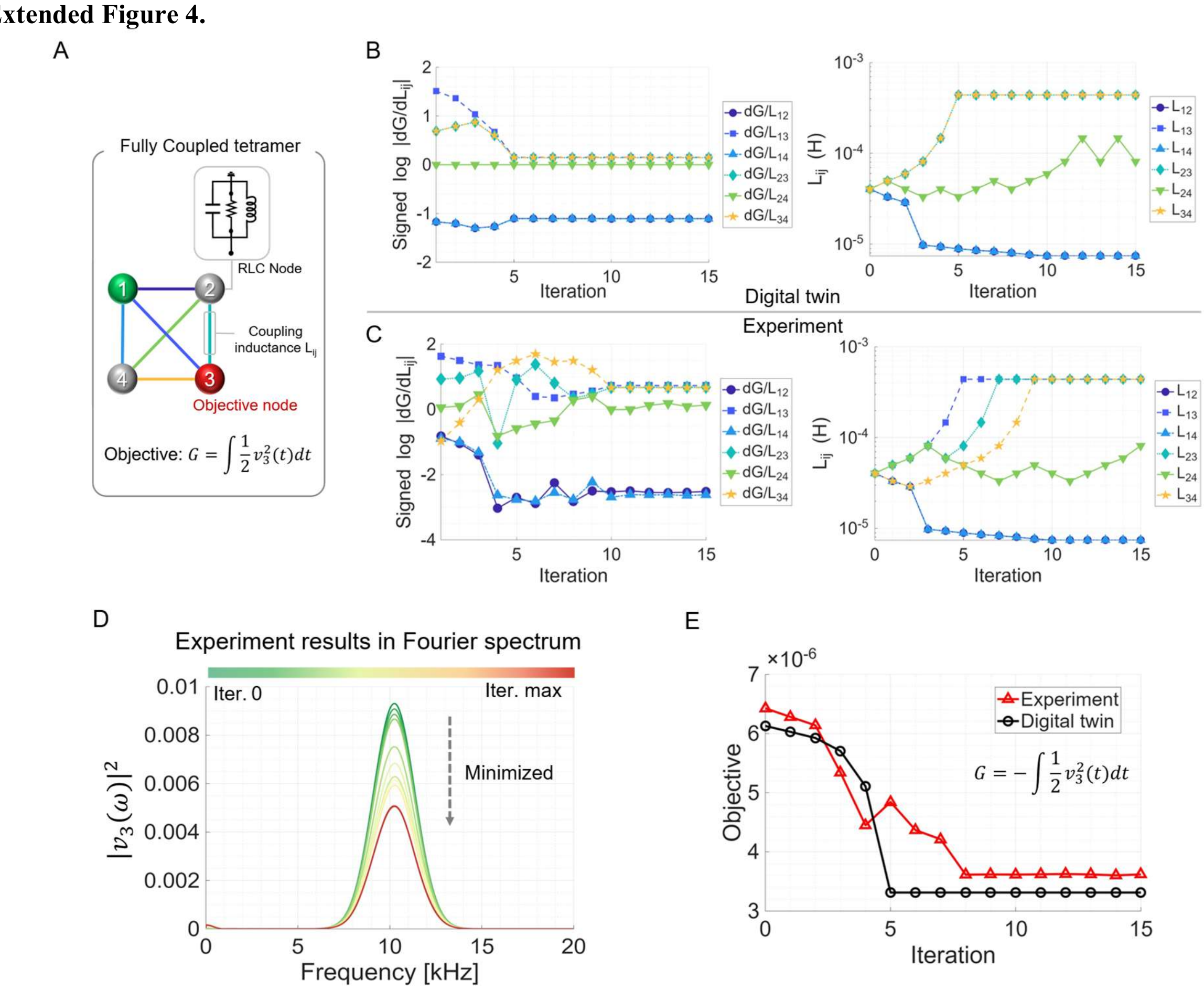

**Extended Figure 4 | Additional optimized results - Broadband minimization**

(A) Schematic of the fully coupled four-node RLC network used for broadband minimization. The objective is to minimize the energy (injected at node 1) at the target node (node 3) by optimizing the coupling inductances. (B) Digital-twin optimization results. The left and right panels show the evolution of the parameter sensitivities and coupling inductances, respectively, during the optimization process. (C) Experimental optimization results. The measured sensitivities and coupling-inductance updates closely follow the trends predicted by the digital twin. The optimized inductance configuration obtained experimentally is consistent with the theoretical prediction, demonstrating accurate gradient estimation and parameter updates in the physical system. (D) Evolution of the measured frequency response during optimization. The spectral power at the target node is progressively reduced near the resonance frequency, indicating successful broadband minimization through iterative parameter updates. (E) Quantitative comparison between digital twin and experiment. The objective-function evolution exhibits similar convergence behavior in both cases, with the optimized objective values approaching the same level.

**Extended Figure 5.**

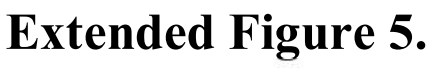

**Extended Figure 5 | Electronic circuit diagrams of fully coupled and chain coupled RLC networks**
(A) Fully coupled 4 RLC network, where all resonator pairs are connected through programmable coupling inductances. Each coupling inductance is digitally controlled using switchable inductive elements, enabling in situ parameter optimization. (B) Chain-coupled 4 RLC network, where only nearest-neighbor resonators are connected through programmable coupling inductances. The same switchable-inductor architecture is used to realize experimentally tunable coupling strengths during optimization.

**Extended Table 1. Component specification**

| Component | Value | Model |
|---|---|---|
| Resistor $R_i$ | 1 kΩ | Stackpole Electronics RNF14FTD1K00 |
| Inductor $L_i$ | 240 μH | Delevan 1537-94J<br>*Chain coupling: 2 parallel connections of $L_i$ |
| Capacitor $C_i$ (chain coupling) | 1000 pF | Kemet C322C102F1G5TA |
| Capacitor $C_i$ (full coupling) | 100 nF | Kemet C322C104K5R5TA |
| Coupling inductor 1 $L_{\text{coupling}}^{(1)}$ | 10 μH | Bourns 78F100J-TR-RC<br>*Chain coupling: 2 serial connections of $L_{\text{coupling}}^{(1)}$ |
| Coupling inductor 2 $L_{\text{coupling}}^{(2)}$ | 56 μH | Bourns 78F560J-RC |
| Coupling inductor 3 $L_{\text{coupling}}^{(3)}$ | 100 μH | Bourns 78F101J-TR-RC |
| Coupling inductor 4 $L_{\text{coupling}}^{(4)}$ | 220 μH | Bourns 78F221J-RC |
| Terminating load $R_T$ | 100 Ω | Vishay Dale CMF55100R00FHEB |
| Single pole single throw switch $S_{ij}$, $S_F$, $S_A$ | N/A | Analog Devices ADG1612BRUZ-REEL7 |
| Microcontroller | N/A | Arduino Nano A000005 |

**Extended Table 2. Device specification**

| Device | Model |
|---|---|
| Arbitrary waveform generator | B&K Precision 4064B - 120 MHz Dual Channel Function/Arbitrary Waveform Generator |
| Oscilloscope | Tektronix DPO2014 Digital Phosphor Oscilloscope, 100 MHz, 4 Ch., 1 GS/s |
| DC Power supply | Rigol DP932E - Three Channel Programmable DC Power Supply |
| Breadboard | REXQualis Small Breadboard 400 Point Solderless Breadboards |